\documentstyle[12pt,fleqn,elsevier,epsfig]{article}
\renewcommand{\baselinestretch}{1.2}

\begin{document}
\setcounter{footnote}{0}
\renewcommand{\thefootnote}{\alph{footnote}}

\vspace{2cm}
\begin{center}
{  \Large \bf 
Solution to the ion feedback problem\\
in Hybrid Photon Detectors and Photo 
Multiplier Tubes\footnote{accepted for publication in Nucl. Inst. Meth. A}
\\[0.5cm]}

\noindent
Daniel~Ferenc$^{1,2}$\footnote{e-mail: Daniel.Ferenc@cern.ch},
Dario~Hrupec$^{1}$ and \\
e-mail: Daniel.Ferenc@cern.ch
Eckart~Lorenz$^{3}$


\noindent
$^1$Rudjer Bo\v{s}kovi\'c Institute, Zagreb, Croatia,\\
$^2$ CERN, Geneva, Switzerland,\\
$^3$Max--Planck--Institut f\"ur Physik, M\"unchen,
Germany

\end{center}

\vspace{2cm}

\renewcommand{\baselinestretch}{0.0}
\small\normalsize

\begin{center}
{\bf Abstract}
\end{center}

A general solution to the positive ion feedback problem
in Hybrid Photon Detectors (HPD), photo multipliers (PM) and other
similar detectors
was found in the insertion of a permanent
electrostatic potential barrier
which prevents drift of positive ions from the anode, or the first dynode,
towards the
photocathode. In this paper we present the method as
applied to
the Intevac HPD.

\vspace{1.2cm}

\renewcommand{\baselinestretch}{1.2}
\small\normalsize
\setcounter{footnote}{0}
\renewcommand{\thefootnote}{\arabic{footnote}}

\section{Introduction}

This work has been motivated primarily by the recent developments
in gamma ray astronomy.
Air Cerenkov Telescopes (ACT) have been considered 
the ultimate instruments
for the ground based detection of
high energy cosmic gamma rays~\cite{ACT-general}.
In order to lower the 
energy threshold for the detection of cosmic gamma rays
down to 20 GeV -- to explore the only unexplored
window in cosmic electromagnetic spectrum (20 GeV to 300 GeV) -- 
one should both 
increase the detector area, and achieve an unprecedented
photon detection with single photon
sensitivity and very high efficiency. Considering photon sensors,
Hybrid Photon Detectors (HPD), 
possibly with high quantum efficiency photocathodes,
currently present 
the most promising solution.
However,
commercial devices have still some 
serious drawbacks and need further improvement.
In particular, it is very important to 
reduce the internal instrumental noise below the limits
imposed by presently available technology, because other
intrinsic sources of noise in imaging
Cerenkov detectors (like the night sky background)
are irreducible.

The presence of positive ions in the vacuum
tube is specially devastating because
the acceleration and subsequent 
dumping of positive ions into the 
photocathode leads both 
to creation of noise
through electrons released,
and 
to a damage of the
photocathode 
~\cite{Razmik,Eckart-NIM}.
In tubes with high vacuum the vast majority of
positive ions do not originate from residual gas, but
from the impact of 
accelerated photoelectrons 
in the surface of the anode (the photodiode and
the surrounding material).
Hydrogen or Oxygen ions from adsorbed water are the
most abundant ions.
Cesium ions are particularly
abundant in devices with in situ photocathode activation procedure.

In this paper we demonstrate,
using 
the Intevac HPD
~\cite{Eckart-NIM} as an example,
how the 
insertion of an electrostatic potential barrier
close to the anode 
solves the ion feedback problem.
Apart from being complete, this 
solution is elegant and easy to
implement.

The paper is structured as follows.
The basic design of the Intevac HPD is described
in Sect.1. Our modification of the 
Intevac design, which includes the potential
barrier, is described in Sect.2.; two different 
actual
designs with equivalent performance 
are presented and compared from a technological
point of view, by carrying out 3-dimensional electron
optics calculation.
In Appendix 1. we discuss for reference an attempt of
Intevac to solve the ion feedback problem by
breaking the field symmetry in the HPD tube.

For all electron optics simulations presented in this 
paper we have used SIMION 3D software~\cite{SIMION}.

\section{Intevac Hybrid Photon Detector}

The basic design of the considered Intevac HPD
~\cite{Eckart-NIM}
comprises a cylindrical vacuum tube, housing a
18 mm diameter photocathode, a set of focussing electrodes,
and a photodiode (PD) (or an avalanche photodiode (APD))
placed at a distance of 39 mm away from the photocathode.
Photoelectrons from the entire photocathode 
are focussed and accelerated
towards a small photodiode, where each electron creates a large number
of electron--hole pairs. The Intevac tube 
is envisaged to be operated at a potential difference
ranging from 5 to 10 kV. The lower end is more suitable
for ACT experiments due to possible harsh atmospheric conditions
at the experimental sites. All simulations presented
in this paper are therefore performed with 5 kV 
potential difference between the photocathode and
the photodiode. The relatively low secondary electron yield
in the PD\footnote{
The number of secondary electrons is N$_e\simeq $(U-U$_0)/3.6$V,
where U$_0$ is the voltage (in Volts) to accelerate an electron 
to an energy sufficient to penetrate the inert layer
above the semiconductor pn structure.
}
may be compensated using an APD with internal
multiplication.

\begin{figure}[t]
\epsfig{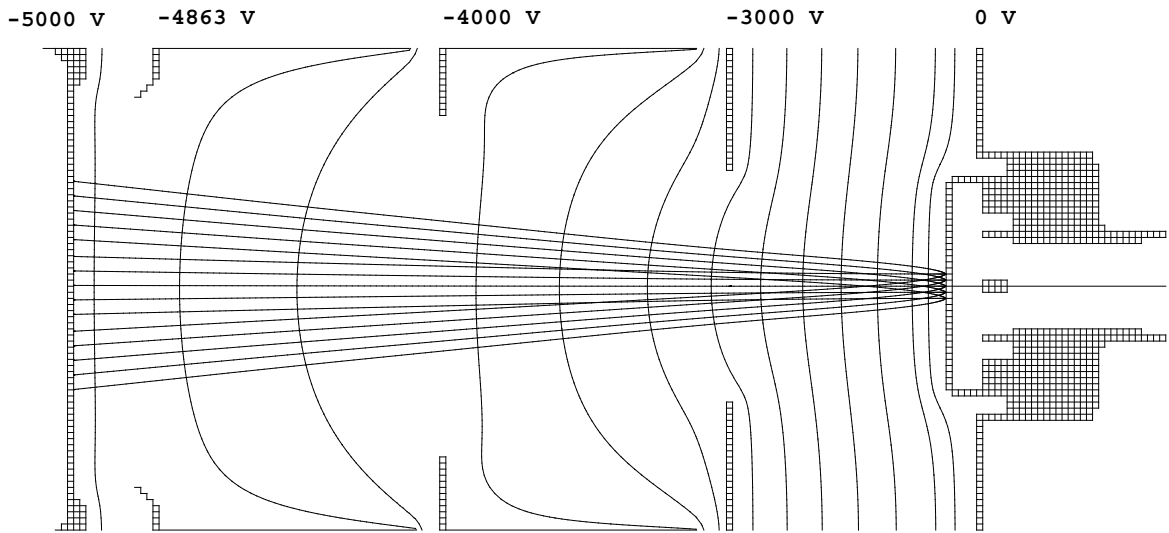}
\caption{
\noindent
Positive ion trajectories in the Intevac HPD. Ions
emerge from the surface of the anode (right) and accelerate
towards the photocathode (left).
}
\label{FPP2}
\end{figure}

The ion feedback problem for the Intevac HPD 
is demonstrated in
Fig.~\ref{FPP2}, where the HPD is shown
with its photocathode on the left, and the anode
complex on the right. Potentials are indicated for
all electrodes, and a set of equipotential
lines is presented (not equidistant). 
Positive ions
of energy E$_{ion}$=15 eV
are launched
from a set of points 
on the anode surface 
with emission angles
+45$^{\circ}$, -45$^{\circ}$ and 0$^{\circ}$.
After being accelerated,
ions hit the photocathode, thus giving rise to the
``ion feedback problem".

Note that the angular and
energetic distributions of positive ions
are, to our best knowledge, unknown. We have worked out a 
scheme how to perform a measurement of those quantities,
using actually a tube designed according to our proposal in this paper, 
but
since the results are not yet available,
we are currently using a rough
estimate that the ions could reach an energy of around 15 eV.
Once the actual energy will be measured,
it will be straightforward to repeat our
simulations and find the optimal potential settings.


\section{Potential barrier -- solution 
to the ion feedback problem}

\begin{figure}[htp]
\epsfig{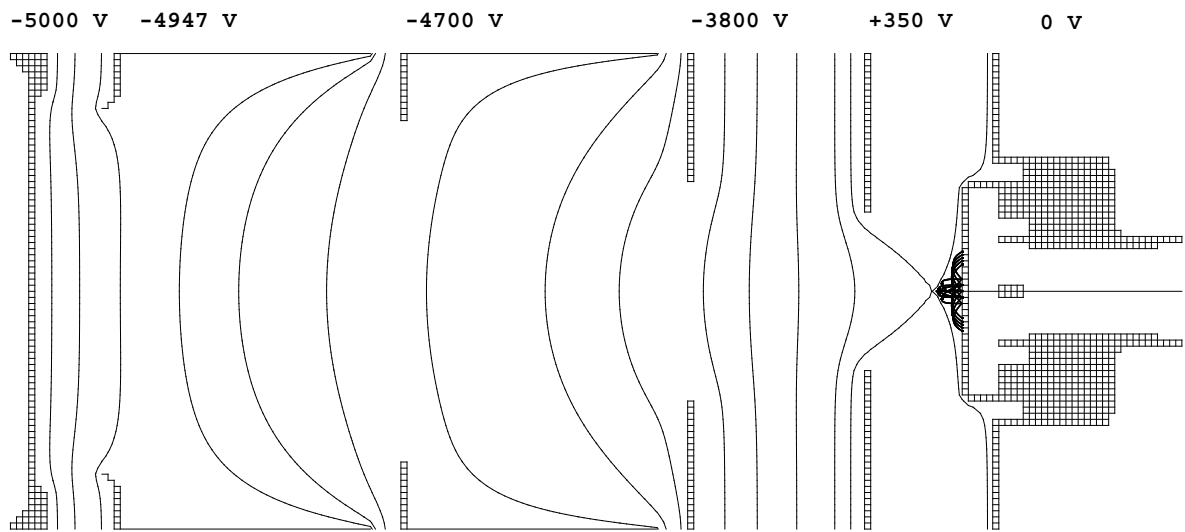}
\caption{
\noindent
HPD design with a flat barrier--electrode at 
potential +350 V. Between the barrier--electrode
and the anode a potential barrier is established,
which repels positive ions emerging from the 
anode surface back towards the anode, see 
trajectories close to the anode surface.
}
\label{FPP3}
\end{figure}

\begin{figure}[hbp]
\epsfig{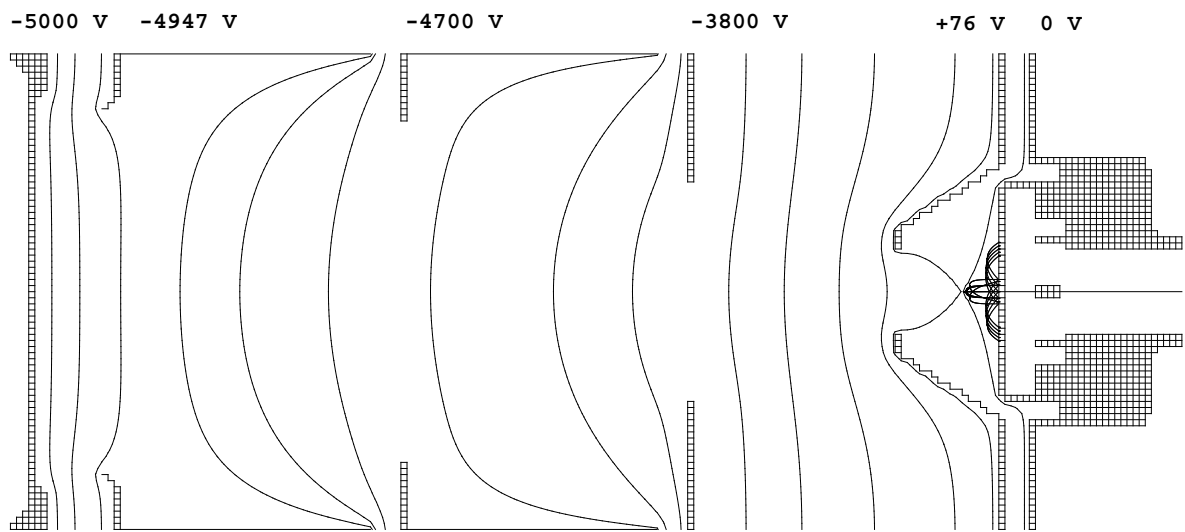}
\caption{
\noindent
The same as in Fig.~\ref{FPP3}, but
for the conical barrier--electrode design.
}
\label{FPP4}
\end{figure}

\begin{figure}[htbp]
\epsfig{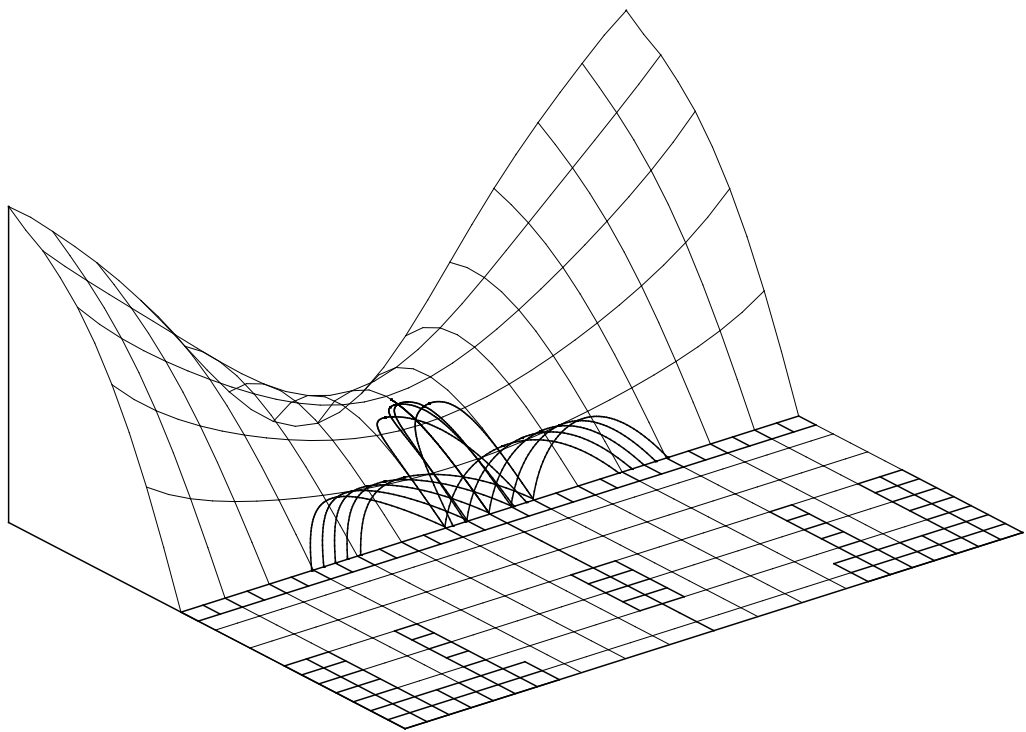}
\caption{
\noindent
Potential distribution in front of the
anode plane of the conical barrier--electrode 
HPD design (Fig.~\ref{FPP4}).
Positive ions of energy E$_{ion}$=15 eV
and emission angles 
+45$^{\circ}$, -45$^{\circ}$ and 0$^{\circ}$ 
start ``climbing" the potential barrier
and get repelled back.
}
\label{FPP7}
\end{figure}

The essential goal of the new electron optics 
is to
prevent positive ions from
leaving the region around the anode 
and heading towards
the photocathode. That goal is accomplished
by means of a permanent electrostatic 
potential barrier
created in front of the anode 
by a cylindrically
symmetric electrode -- the so called ``barrier--electrode" --
kept at a potential somewhat higher
than the anode potential. 
The method preserves cylindric symmetry
of the entire device.
Two different designs
are presented in this paper: (i) 
with a flat barrier-electrode, see Fig.~\ref{FPP3},
and (ii) with a conically shaped barrier--electrode, see Fig.~\ref{FPP4}.
In addition, the potential distribution around the anode plane for
the conical design
is shown in a magnified view in Fig.~\ref{FPP7}.
Trajectories of singly charged positive 
ions are simulated with identical initial conditions
like before.

\begin{figure}[htbp]
\epsfig{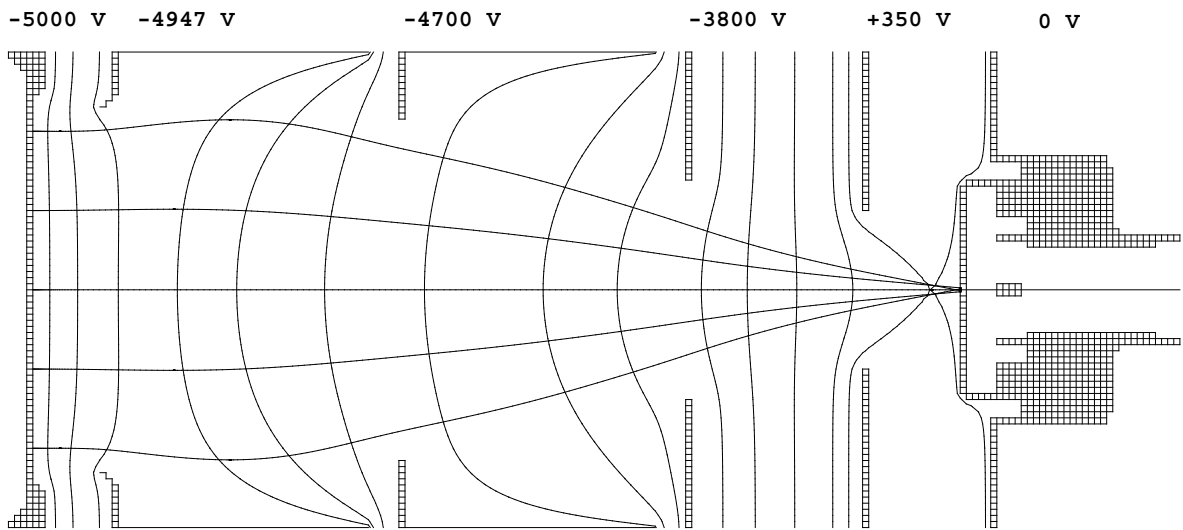}
\caption{
\noindent
Electron focussing with a flat barrier--electrode
HPD.
}
\label{FPP5}
\end{figure}

\begin{figure}[htbp]
\epsfig{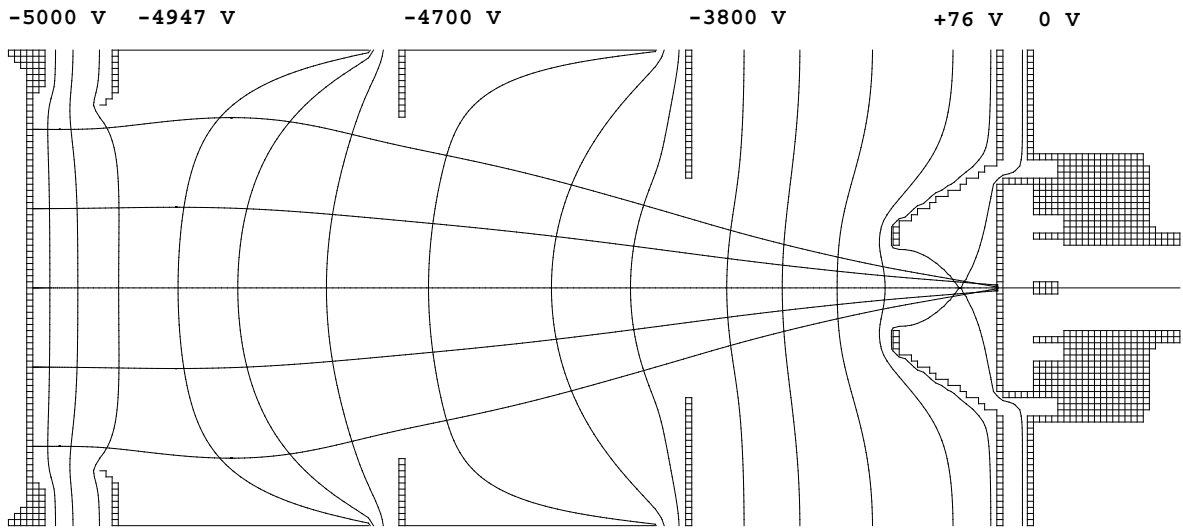}
\caption{
\noindent
Electron focussing with a 
conical barrier--electrode HPD.
}
\label{FPP6}
\end{figure}

The functionality of the barrier--electrode
is simple: it creates a potential barrier
in front of the anode 
which does not
allow ions to penetrate further towards the photocathode, 
solving thus the ion feedback problem.
The design with the conical 
barrier--electrode, Fig.~\ref{FPP4}, offers
certain advantages.
Since the conically shaped electrode more closely
surrounds the anode area, the barrier--electrode potential needed 
to establish the potential barrier of a given height 
(for the examples presented we used V$_{Barrier}$=+15.5 V)
is much lower (V$_{BE}$=+76 V) than needed for 
the flat electrode (V$_{BE}$=+350 V). 
In addition, 
the throughput connection to the voltage 
supply outside the tube appears for
the conical barrier--electrode
at a very safe distance from the throughput of the 
neighboring electrode
of opposite polarity. Further, the conical electrode
may better protect the anode area from cesium
vapor deposition during (and after)
the manufacturing of the 
photocathode, and also captures photoelectrons
back-scattered from the PD.
Both designs are optimised for electron focussing.
Electron trajectories are shown in
Fig.~\ref{FPP5} and
Fig.~\ref{FPP6}, for
the flat and the 
conical barrier--electrode designs, respectively. 
Electron trajectories were simulated 
with the following initial conditions:  emission angle normal to
the photocathode surface, and initial
energy E$_{electron}$=0.4 eV. 
An avalanche photodiode with 1 mm in diameter 
will eventually collect all the electrons, 
even if smearing in
initial electron energy and emission angle
is considered~\cite{Diplomski}.

Let us also mention that some existing HPD and PM tubes
do have already electrodes close to their anodes
or first dynodes - originaly designed for other
purposes - which may be used as our barrier electrodes
with the application of an appropriate potential,
and a redistribution of other potentials
in~the~device~in~order~to~compensate~for~the~likely~change~in~electron~focusing.

The stability
of the potential on the barrier--electrode which
is required
for stable electron focussing is
not a critical issue
-- variations of even 10\% on the potential will leave
the electron focussing essentially 
unchanged~\cite{Diplomski}.
The most common 
voltage supply may be therefore used
to bias the barrier--electrode.

\newpage
\section{Summary}

A general solution to the positive ion feedback problem
in hybrid photon detectors, photomultipliers and other
similar detectors
was found in the creation of an
electrostatic potential barrier in front of
the anode in HPD's or the fitst dynode in PM's. 
By that means positive ions originating from the 
anode surface - released in impacts of photoelectrons -
are not able to penetrate the potential barrier, and therefore
cannot
get accelerated 
towards the photocathode.

In this paper we presented the
application of the potential barrier method to
the Intevac HPD.
The method is generally applicable, and we have 
applied it recently also in designs of some
other devices~\cite{LHC-B}.

\newpage

\section{Appendix 1. Intevac ion deflector}

\begin{figure}[b]
\epsfig{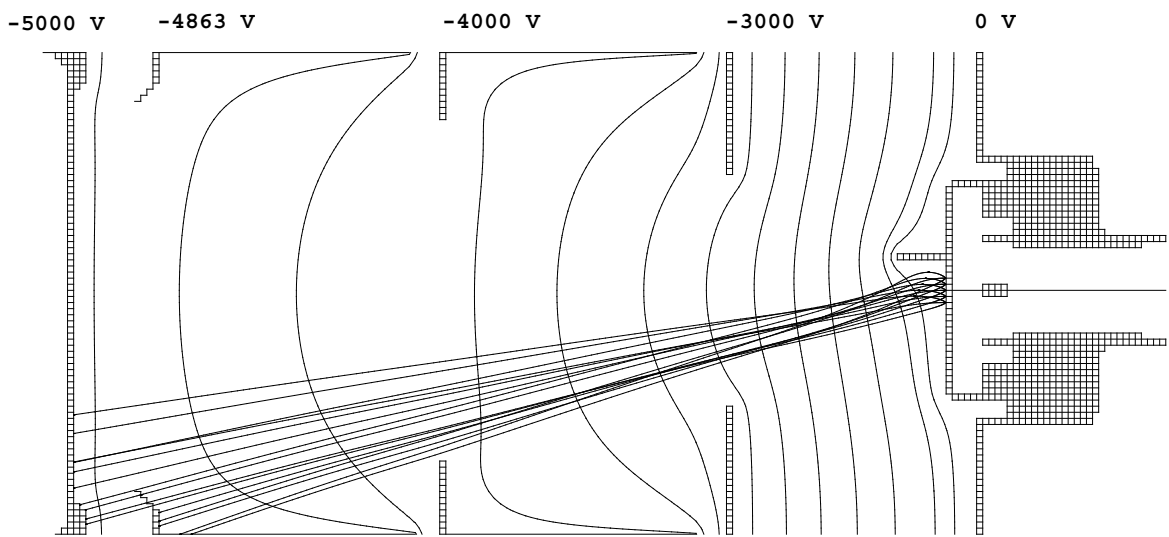}
\caption{
\noindent
Intevac deflector modifies the potential distribution
in such a way that positive ions find their
way to the opposite side of the tube, but they all
hit, with sizeable energies already, either 
the photocathode or other parts of
the device.
}
\label{FPP1}
\end{figure}

In an attempt to solve the ion feedback
problem,
Intevac 
introduced~\cite{Eckart-NIM} a method based on 
broken cylindric symmetry of the
field in the tube -- an idea 
essentially different from ours, because
there is no attempt to block the propagation of ions
towards the photocathode,
but only to deflect their trajectories to one side of
the tube.
The symmetry breaking is
done by inserting an electrode
on the anode potential, 
placed just on one side of the diode~\cite{Eckart-NIM},
the advantage being that no additional positive voltage source is 
needed. Note that the high energy photoelectrons
are barely deflected by the deflector electrode. This deflection
was compensated by displacing the pickup diode by small distance.
A reproduction of the Intevac deflector solution is
presented in Fig.~\ref{FPP1}. The same set of 
ions is simulated like above.

Although positive ions 
become deflected to the opposite side of the
deflector electrode, a considerable fraction
of ions still hit the photocathode and the remaining ones
hit the electrodes or the tube walls, eventually
releasing additional electrons and ions into the 
chamber vacuum -- a result far from being satisfactory.


\newpage
\noindent
\begin{thebibliography}{[99]}

\bibitem{ACT-general} ~J.A. Barrio et al., The MAGIC Telescope Design Study, 
Max-Planck-Institut f\"ur Physik, M\"unchen, 
Internal report
MPI-PhE/98-5, 1998.
\bibitem{Razmik} ~R. Mirzoyan, 
E. Lorenz, D. Petry, and C. Prosch, 
Nucl. Inst. Meth. A387 (1997) 74.
\bibitem{Eckart-NIM} ~S. Bradbury,
R. Mirzoyan, J. Gebauer, E. Feigl, 
and E. Lorenz, Nucl. Inst. Meth. A387 (1997) 45.
\bibitem{SIMION} ~"SIMION 3D Version 6.0" by David A. Dahl 
43rd ASMS Conference on Mass Spectrometry and Allied Topics, 
May 21-26 1995, Atlanta, Georgia, pg 717.
\bibitem{Diplomski} ~D. Hrupec, Diploma thesis, Rudjer Bo\v{s}kovi\'{c}
Institute, and University of Zagreb, 1997.
\bibitem{LHC-B} ~D. Ferenc,``New developments in Hybrid Photon Detectors",
Presented at the Conference
New Detectors Erice,
36th Workshop of the INFN Eloisatron Project,
Erice, Trapani, Sicily,
November 1-7, 1997. To appear in the proceedings,
World Scientific.


\end{thebibliography}
\end{document}